\documentclass[aps,12pt,msam,msbm]{iopart}

\usepackage{graphicx}
\usepackage{amssymb}
\usepackage{iopams}
\usepackage{wasysym}

\begin{document}

\newcommand{\nn}[1]{{\langle #1 \rangle}}
\newcommand{\nnn}[1]{{\langle \langle #1 \rangle \rangle}}
\newcommand{\hc}{\mathrm{h.c.}}
\newcommand{\bra}[1]{{\langle #1 |}}
\newcommand{\ket}[1]{{| #1 \rangle}}

\title[Three-Body Interactions on the Honeycomb Lattice]{Polar Molecules with Three-Body Interactions on the Honeycomb Lattice}
\author{Lars Bonnes, Hanspeter B\"uchler,  and Stefan Wessel}
\address{Institut f\"ur Theoretische Physik III, Universit\"at Stuttgart, Pfaffenwaldring 57, D-70550 Stuttgart, Germany}
\ead{lars@itp3.physik.uni-stuttgart.de}
\date{\today}

\begin{abstract}

We study the phase diagram of 
ultra-cold bosonic polar molecules loaded on a two-dimensional optical lattice of hexagonal symmetry  controlled by external   electric and microwave fields.
Following a recent proposal 
in Nature Physics \textbf{3}, 726 (2007), 
such a system is described by 
an extended Bose-Hubbard model of hard-core bosons, that includes both extended two- and  three-body repulsions.
Using quantum Monte-Carlo simulations, 
exact finite cluster calculations and the tensor network renormalization group, we explore the  rich phase diagram of this system,  
resulting from the strongly competing nature of the three-body repulsions on the honeycomb lattice.
Already in the classical limit, they induce complex solid states with large unit cells and macroscopic ground 
state degeneracies at different fractional lattice fillings. For the quantum regime, 
we obtain effective descriptions of the various phases in terms of emerging valence bond crystal states and quantum dimer models.
Furthermore, we access 
the experimentally relevant parameter regime, and determine the stability of the crystalline phases towards strong two-body interactions.

\end{abstract}


\maketitle

\section{Introduction}
The interplay of competing interactions and quantum fluctuations is  known to allow for interesting  phases to emerge in  many-body quantum systems. This route towards novel phases of  matter has been explored intensively in recent years,  in particular in the field of low-dimensional quantum magnetism~\cite{schollwoeck04}, and  in the 
context of ultra-cold quantum gases on optical lattices, where one gains a high degree of control of the  interaction strength~\cite{bloch08}.
The dominant inter-particle potentials in such systems are typically described by two-body interaction and exchange terms.
It  appears fruitful to explore also realistic set-ups of many-body quantum systems that are dominated -- via engineered  interaction potentials -- by  multi-body interaction terms  of e.g. three-particle type.
Indeed, recently, polar molecules have been proposed as  promising 
candidates towards realizing such many-body systems.  
Driven by  significant progress towards producing degenerate gases of polar molecules~\cite{sage05,ni08,deiglmayr08,lang08,danzl08}, various proposals have been put forward, how to drive ultra-cold polar molecules into regimes of strong many-body interaction effects~\cite{kotochigova06,micheli06,wang06a,buechler07,lahaye09}.
In a recent work~\cite{buechler07} an effective interaction potential
was derived for polar molecules on an optical lattice in the  presence of static electric and microwave fields. It was found, that upon appropriately tuning the external fields, the  interactions between the polar molecules
become  characterized by extended strong two- as well as three-body interactions. 

Here, we derive the ground state phase diagram for these systems with dominating three-body interaction 
using  quantum Monte-Carlo simulations, exact finite cluster calculations and the tensor network renormalization 
group. We find the presence of solid structures for unconventional filling factors with unit cells much larger than 
the period of the underlying lattice,  and a macroscopic ground state degeneracy in the classical limit. 
In the quantum regime, these degeneracies of the low energy sector are lifted giving rise to valence bond 
crystal states.

\begin{figure}
\begin{center}
\includegraphics[width=250pt]{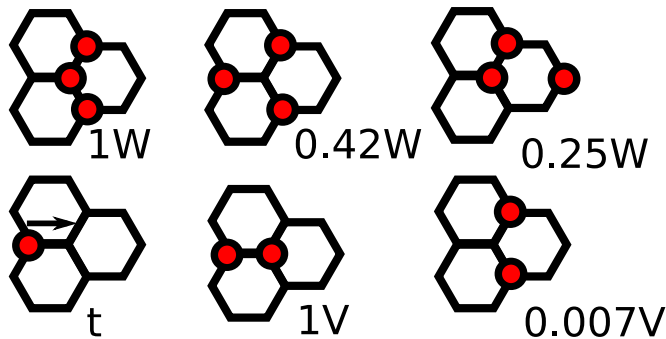}
\caption{
Top row: Leading three-body repulsion terms and their relative strengths on the honeycomb lattice. 
Bottom row: Illustration of the nearest neighbor hopping term, and the relative strengths of the
leading two-body repulsion terms on the honeycomb lattice.
}
\end{center}
\label{fig:interactions}
\end{figure}

The interaction potential for polar molecules within an opitcal lattice in the  parameter regime, where three-body and two-body interactions 
are present takes the form
\begin{equation}
V_\mathrm{eff} = \frac{1}{2} \sum_{ij} V_{ij} n_i n_j + \frac{1}{6} \sum_{ijk} W_{ijk} n_i n_j n_k,
\end{equation}
with $V_{ij}=V/r^6_{ij}$ and $W_{ijk}=W/(r^3_{ij} r^3_{jk} + \mathrm{perm.})$. Here, $r_{ij}$ denotes the spatial separation between  particles on lattice sites $i$ and $j$, and $n_i$  the local density at site $i$.   The the two-body interaction $V$ and the three-body interaction $W$
 can be tuned to be of similar strength, $V\gtrsim W$~\cite{buechler07}.
In the following, we consider in particular the case of bosonic polar molecules. In this case, the suppressed tunneling of a particle to an already occupied site needs to be accounted for by a hard-core constraint on the bosonic occupations~\cite{buechler07}. 
Since the extended interactions decay rapidly with the inter-particle separations (c.f. \Fref{fig:interactions}),  we  truncate the interactions after the leading terms, resulting thus in an extended Bose-Hubbard model with two- and three-body nearest neighbour interactions only,
\begin{equation}
H =	- t \sum_\nn{ij} \left( b_i^\dagger b_j + \hc\right)
	+ V \sum_\nn{ij} n_i n_j 
	+ W \sum_\nn{ijk} n_i n_j n_k
	-\mu \sum_{i} n_i.
	\label{eq:main_ham}
\end{equation}
Here, 
$b_i$ and $b_i^\dagger$ denote boson annihilation and creation operators respectively, and the local density operator $n_i=b_i^\dagger b_i$ has eigenvalues $0$ and $1$ in the hard-core limit. Furthermore, $t$ denotes the nearest-neighbor hopping matrix element, and $\mu$ is a chemical potential, allowing to control the filling (i.e. the density) of the system between $n=0$ (empty) and $n=1$ (full). 

In  previous works, models of hard-core bosons with extended two- and three-body interactions as effective models of ultra-cold polar molecules were studied for the case of a one-dimensional optical lattice~\cite{capogrossosansone09} and the two-dimensional square lattice geometry~\cite{schmidt08}. In the one-dimensional case, an incompressible phase at a filling $n=2/3$ was established for dominant three-body interactions, stabilizing both change-density wave (CDW) and bond-order wave (BOW) long-ranged correlations, apart from  conventional CDW phases appearing at half-filling ($n=1/2$) in the presence of two-body interactions. On the square lattice, several solid phases at fractional fillings as well as supersolid phases were found in a semi-classical approximation, some of which could also be verified by full numerical simulations.

Here, we extend such systematic explorations to the case of the honeycomb lattice, where  nearest-neighbour three-body repulsions lead to characteristic effects  of strong frustrations.
In order to explore the  physics of this system, we used a combination of quantum Monte Carlo (QMC) simulations based on the generalized directed loop algorithm  within the stochastic series expansion (SSE) representation~\cite{sandvik99b,syljuasen02,alet05}, as well as  exact finite cluster calculations and the tensor network renormalization group approach~\cite{levin07} for the classical ($t=0$) limit of the above Hamiltonian, as detailed below. 
In the following Section 2, we  discuss the results of our calculations on the ground-state phase diagram  in the absence of  two-body interactions (i.e. for $V=0$), thus focusing on the main aspects of  three-body repulsions on the honeycomb lattice. We observe states with large degeneracies in the classical limit, and 
the emergence of complex valence bond crystal (VBC) phases, to be described in detail below. 
In Section 3, we discuss the behaviour of the system as we perturb it by a finite two-body repulsion $V$, and   explore the experimentally relevant parameter regime where $V\gtrsim W$~\cite{buechler07}. We examine the stability range of the 
VBC phases, and find that a  cascade of incompressible solid phases steams from the competition between  two- and three-body repulsion terms.

\section{Three-Body Interactions}
\begin{figure}
\centering
\includegraphics[width=300pt]{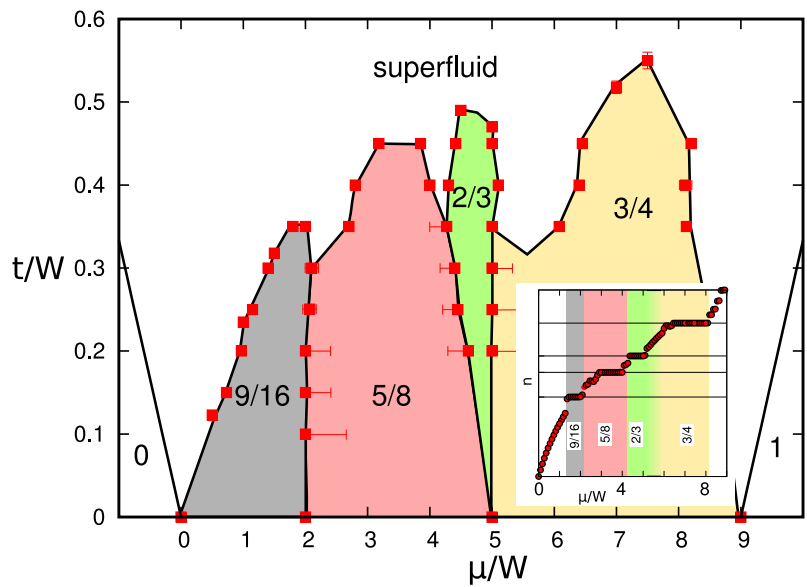}
\caption{
Ground-state phase diagram of hard-core bosons on the honeycomb lattice for $V=0$ in terms of $\mu/W$ and $t/W$.  The incompressible phases at fillings $n=9/16$, $5/8$, $2/3$ and $3/4$ are labeled by $n$ and underlayed by different colors. Uncertainties on the estimated phase boundaries are indicated by error bars.  
The inset shows the density $n$ as a function of the chemical potential $\mu/W$ for fixed $t/W=0.3$, linear system size $L=12$ and an inverse temperature of $\beta=20W$.
}
\label{fig:phase}
\end{figure}

\begin{figure}
\centering
\includegraphics[width=300pt]{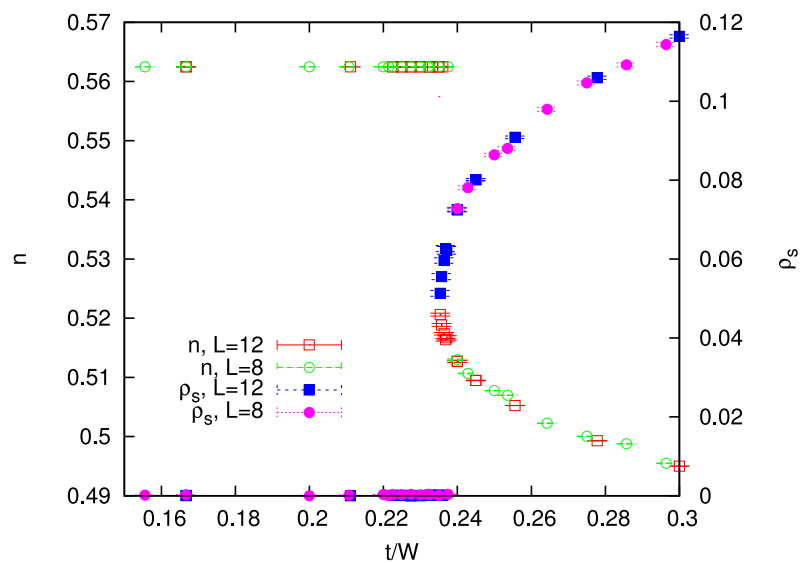}
\caption{
Behavior of the density $n$ and the superfluid density $\rho_s$ near the quantum melting transition between the $n=9/16$ incompressible phase and the superfluid phase at $\mu/W=1$ and $V=0$ taken at $\beta=100$.}
\label{fig:melting}
\end{figure}

In this section, we focus on the case $V=0$ in order to explore explicitly the effects of three-body repulsions on the honeycomb lattice. The ground state phase diagram in \Fref{fig:phase} summarizes the results from our analysis. It exhibits at low values of $t/W$ a variety of incompressible phases of different (unconventional) fillings  $n=9/16$, $5/8$ $( =10/16)$, $2/3$ and $3/4$ $( =12/16)$.
Due to the incompressible nature of these incompressible phases, they lead to finitely extended plateaus in the $\mu$-dependence of the density $n$, as shown e.g. for fixed $t/W=0.3$ in the inset of \Fref{fig:phase}. The actual nature of these 
 phases and the quantum phase transitions between them, will be discussed below. 
For larger values of $t/W\gtrsim 0.4$, the system is eventually driven by its kinetic energy from 
these solid phases via first-order quantum melting transitions into a uniform superfluid phase with a finite superfluid density $\rho_s$. 
In the QMC simulations, the superfluid density is obtained 
as $\rho_s=\nn{w^2}/(\beta t)$ 
from measuring the bosonic winding number $w$ fluctuations in the standard way~\cite{pollock87} (here, $\beta=1/T$ ($k_B=1$) denotes the inverse temperature). The first-order nature of the melting transitions follows from pronounced  jumps that are observed upon crossing the transition lines in both the density and the superfluid density.
As a typical example, we show in \Fref{fig:melting} the behavior of $n$ and $\rho_s$ near the quantum melting transition between the  $n=9/16$ incompressible phase and the superfluid phase at $\mu/W=1$. The first-order nature of the quantum melting transitions is indeed expected, since (as shown below) the incompressible phases break the space group symmetry, whereas in the uniform superfluid  $U(1)$ symmetry breaking occurs at $T=0$. 

\subsection{Quantum Monte Carlo}

In order to perform  the SSE QMC simulations for the current model, we employed a cluster decomposition of the Hamiltonian in terms of trimers of nearest-neighbor sites, such that each cluster carries  one of the three-body interaction terms. The addition of the nearest neighbor two-body repulsions $V$ (in Section 3), 
proceeds  in this decomposition scheme as well, with each two-body term being shared by four such trimers. In the SSE directed loop construction, we thus used a doubly-linked list of 6-leg vertices. While the algorithm performed well for large values of the hopping $t$, we were not able to reach significantly below $t/W\approx 0.1$ due to the dynamical freezing in the Monte Carlo configurations, once the competing diagonal interaction terms dominate the Hamiltonian (which is a general issue of the SSE for Hamiltonians dominated by large diagonal terms).   
In terms of the simulation temperature $T$, we find that an inverse temperature $\beta=1/T=20-50W$ ($k_B=1$) provided an optimal trade-off between the finite temperature incoherence and the algorithmic performance (the SSE algorithmic cost  scales linearly in $\beta$). Furthermore, in order to be commensurate with all superstructures identified in the 
incompressible phases, the linear system size $L$ is required to be an integer multiple of $6$ (the total number of lattice sites being $N=2L^2$, as the honeycomb lattice contains two sites per unit cell, forming the two sublattices $A$ and $B$). 
Within the above temperature regime, we were able to simulate systems  up to $L=24$, in some cases up to $L=36$ in linear extend. 
In  phases with large superstructures  -- described below
--  we find that the autocorrelation times of the bosonic structures increase such that the algorithm is not able to tunnel between different realizations of the ordering pattern even within several $10^6$ QMC sweeps, but resides in one particular sector of the ground-state manifold (which however varies upon performing independent runs with different random-number streams for a given set of model parameters). During a first stage of the thermalization process,
we  annealed the system in a cyclic way by heating it up and cooling it down slowly such that it is able to relax globally into one of the equivalent ground-states. A fixed temperature thermalization was then performed in the second stage of the thermalization process.
This annealing approach yield  better performance than parallel tempering over an extended temperature ranges, since we are mainly interested in the ground-state phase diagram of the system, where $\beta$ is large.


We furthermore employed quantum parallel tempering \cite{sengputa02} (in $\mu$ or $t$) at fixed low temperatures in certain regions of the phase diagram, in particular in order to study  the quantum phase transitions between neighboring VBC phases. We discuss these  aspects in more detail in Section 2.3.
Our special analysis of the model in the classical limit (i.e. for $t=0$) is presented in Section 2.4. There, we also assess  the applicability of the tensor network renormalization group approach for classical systems to the current  three-body repulsive model.  
In the following Section 2.2, we  discuss in detail the nature of the incompressible phases that appear in \Fref{fig:phase}, and how they are characterized from our numerical analysis. 

\subsection{Incompressible Phases}

\paragraph{The $n=9/16$ VBC Phase:}

\begin{figure}
\centering
\includegraphics[width=300pt]{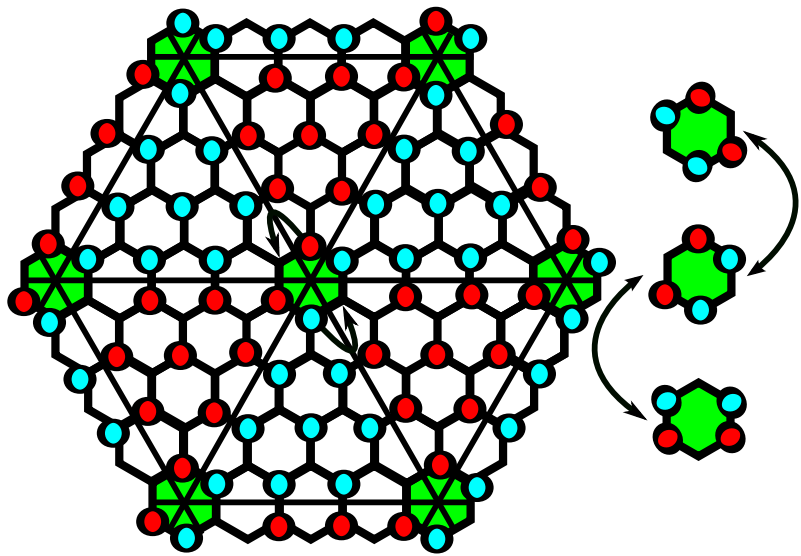}
\caption{
Illustration of the classical configuration in the $n=9/16$ plateau (left panel). The bosons are shown in two different colors, in order to underline the different checkerboard patterns within neighboring triangles. 
The green hexagons illustrate those plaquettes, where the bosons are allowed to change positions 
as illustrated in the right panel, without changing the potential energy. For finite hopping $t>0$, this leads to resonances within these hexagonal plaquettes (right panel).
}
\label{fig:9_16}
\end{figure}

The first density plateau that is encountered upon filling the lattice has a density $n=9/16$ and extends between $0\le \mu/W \le 2$ in the classical limit $(t=0)$. It has an potential energy per lattice site (equal to the internal energy for $t=0$) of $E_{pot}^{(9/16)}=-9/16\mu$, and corresponds to the closest packing of hard-core bosons on the honeycomb lattice without introducing any three-body repulsions. 
From geometrical considerations in the classical limit, 
this bosonic structure is obtained by covering the lattice with equilateral triangles with side length $4\sqrt{2} a$ (where $a$ is the distance between two lattice sites), each covering 16 lattice sites. These triangles are filled by nine bosons in a staggered (checkerboard) arrangement in order to obtain the overall filling $9/16$. Neighboring triangles differ in the placement of the checkerboard pattern. This leads to domain walls 
along the edges of the triangles, where pairs of bosons reside. For $V=0$, this does however not lead to a potential energy penalty. 
For an illustration of this structure, see the left panel of \Fref{fig:9_16}.
This structure allows for a denser packing  of the particles (and thus a higher filling) than  the overall checkerboard state of filling $n=1/2$, without introducing any three-body energy terms.
On every forth hexagon (such as the hexagons indicated in the left panel of \Fref{fig:9_16}), six triangles share a  common corner.
In the classical limit, the energy of the system remains unchanged, if two particles change their positions along such a hexagon by an angle of $2\pi/6$, as indicated in \Fref{fig:9_16}. This local move results in a classical ground state 
degeneracy $W=3^{N/32}$. Thus, the ground state entropy is extensive, and the entropy per site is $S/N=(\ln W)/N=\log(3)/32 \approx 0.034$ in the thermodynamic limit.  

For finite values of $t$, 
the local moves allowed in the classical configurations lead to  
resonances on the  hexagonal plaquettes,  corresponding to second-order hopping processes of the bosons, effectively rotating a hexagon by an angle of $2\pi/3$, as illustrated in the right panel of \Fref{fig:9_16}. The system is able this way to gain kinetic energy from these tunneling processes. Such resonances also provide the dominant quantum fluctuations on this density plateau, and stabilize a VBC phase with a superstructure of checkerboarded triangles, linked by the resonating hexagons. 
Through an 
order-by-disorder 
effect, the quantum dynamics thus 
selects the ground-state to be a coherent superposition of the local resonance states. 

We can indeed identify these peculiar features of the $9/16$ plateau  phase from the QMC simulations. 
\begin{figure}
\centering
\includegraphics[width=300pt]{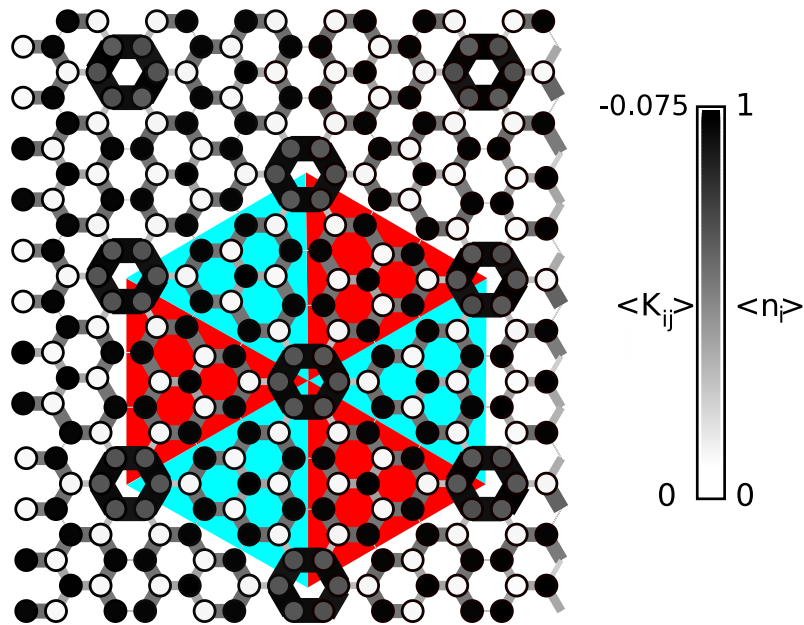}
\caption{QMC data of the local density $\nn{n_i}$ (shading) and the kinetic energy density along the nearest-neighbor bonds $\langle K_ {ij}\rangle$ (line thickness and shading) for bosons on the honeycomb lattice 
in the $n=9/16$ VBC phase
at $V=0$, $t/W=0.2$, $\mu/W=1$, and a system with $L=12$ at $\beta=20$.
For clarity, the equilateral triangles, together forming the unit cell of the VBC superstructure, are indicated as well. 
}
\label{fig:9_16__MC__filling}
\end{figure}
In \Fref{fig:9_16__MC__filling}, we show the local density $\nn{n_i}$, along with the kinetic energy density per bond $\langle K_{ij}\rangle $, where $K_{ij}= b_i^\dagger b_j+b_j^\dagger b_i$
and sites $i$ and $j$ belong to a nearest neighbor bond on the honeycomb lattice, for a representative point within the $n=9/16$ phase.
We find 
the local  density is close to $\nn{n_i}=1$ and $0$,  thus exhibiting only few fluctuations, in a staggered (checkerboard) pattern within triangular structures.
On the other hand, the density around a subset of the hexagons -- those, where the triangular checkerboard patterns meet -- is 
within an intermediate range, and it is along the bonds of these hexagons, where the
kinetic energy is mainly located. Both observations indicate a residual density dynamics in this incompressible phase,  located along these hexagons. 
\begin{figure}
\centering
\includegraphics[width=230pt]{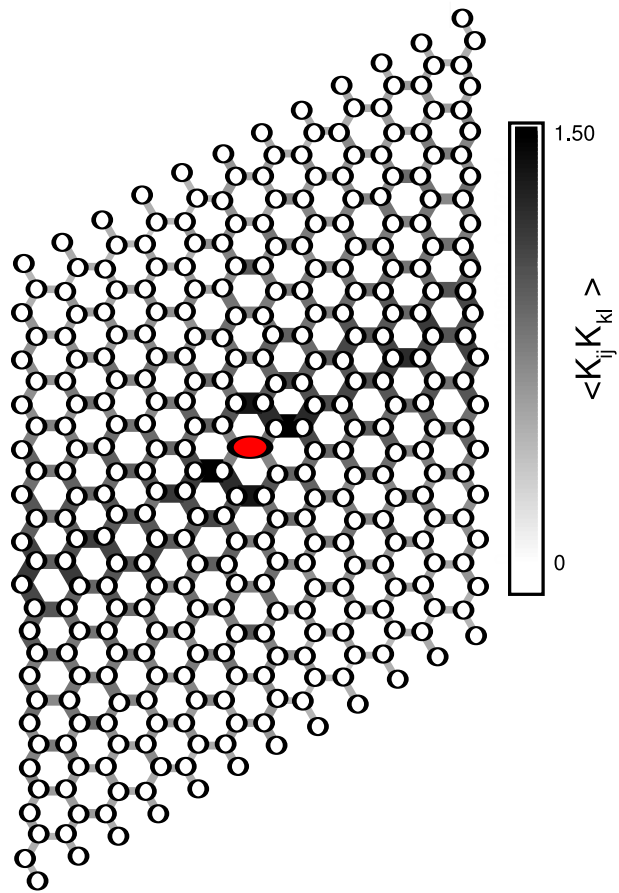}
\caption{QMC data for the bond-bond correlations in the kinetic energy $\langle K_{ij} K_{kl} \rangle$ along the nearest-neighbor bonds for bosons on the honeycomb lattice 
in the $n=9/16$ VBC phase
at $V=0$, $t/W=0.2$, $\mu/W=1$, and a system with $L=12$ at $\beta=20$. The reference bond $\nn{ij}$ is indicated by the red ellipse. }
\label{fig:9_16__MC__kinetic}
\end{figure}
The resonant nature of the corresponding hopping events is  reflected 
in the bond-bond correlation function $\langle K_{ij} K_{kl} \rangle$, shown in \Fref{fig:9_16__MC__kinetic}. 
We identify the main contribution from the hexagonal resonances,
which lead to the enhanced bond-bond correlation between the reference bond (marked by an ellipse) and the bonds atop and below the reference bond in \Fref{fig:9_16__MC__kinetic}.
Furthermore, correlations of the same strength are visible between the reference bond and its next-nearest neighbour bonds  to the left and right.
These result from  residual quantum fluctuations that lead to the finite kinetic energy distribution inside the checkerboarded triangular structures in \Fref{fig:9_16__MC__filling}. We do not observe long ranged bond-bond correlations, thus the hexagonal resonances evolve essentially independently of each other.

\begin{figure}
\centering
\includegraphics[width=400pt]{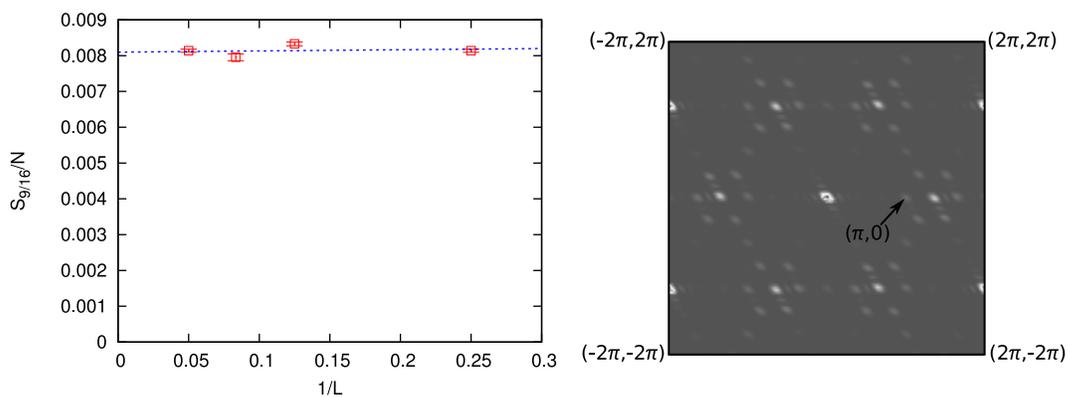}
\caption{Finite size scaling 
of the density structure factor $S_{9/16}$ for the $n=9/16$ phase of hard-core bosons on the honeycomb lattice at 
$V=0$, $t/W=0.1$, and $\mu/W=1$ (right panel). The left panel shows the positions of the peaks in the density structure factor in momentum space, with the peak at $\vec{q}_0=(\pi,0)$ indicated by the arrow.
}
\label{fig:9_16peak}
\end{figure}
The left panel of \Fref{fig:9_16peak} shows the density structure factor for the $9/16$ VBC structure, $S(\vec{q})=1/N \sum_{ij} n(\vec{x}_i) n(\vec{x}_j)\: e^{i \vec{q} (\vec{x_i} - \vec{x_j})}$. Here, $n_a(\vec{x}_i)$ denotes 
the local density operator at position $\vec{x}_i$.
A 6-fold  structure is identified, with one of the equivalent peaks positioned at $\vec{q}_0=(\pi,0)$. This structure relates to the superstructure of equilateral triangles and thus provides a characteristic feature in the density distribution of this phase. 
The right panel of \Fref{fig:9_16peak} shows a finite size scaling analysis of  QMC data for the structure factor 
$S_{9/16}=\langle S(\vec{q}_0)\rangle$ 
at this characteristic wavevector $\vec{q}_0$ for a system  within the $9/16$ VBC phase. 
We find that $S_{9/16}/L^2$ indeed extrapolates in the thermodynamic limit ($N\rightarrow\infty$) to a finite value, verifying the presence of long-range density order in the bosonic structure of the $9/16$ VBC phase.

\paragraph{The $n=5/8$ VBC Phase:}

The next density plateau encountered upon further increasing the chemical potential  appears between $2< \mu/W \le 5$ in the classical limit, and has a filling of $n=5/8$. The classical potential energy equals  $E_{pot}^{(5/8)}=-5/8\mu +W/8$ in this phase.
As for the $9/16$ plateau, we obtain a VBC structure in the quantum regime. However, its effective description is more involved. 
\begin{figure}
\centering
\includegraphics[width=200pt]{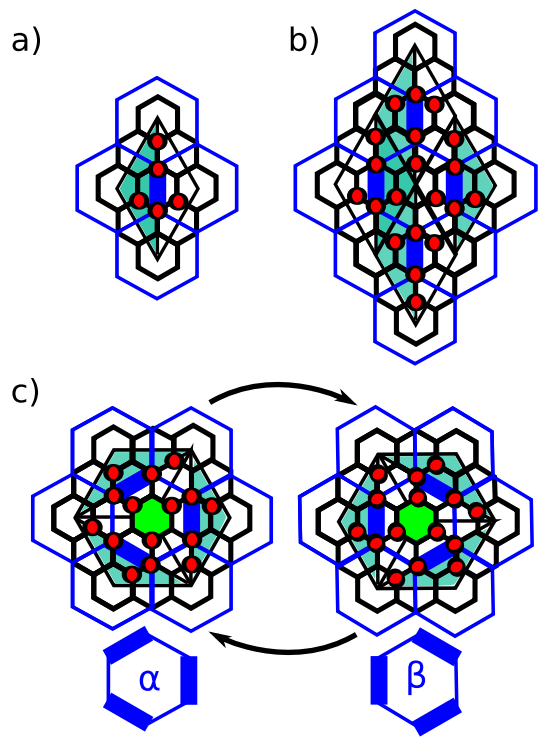}
\caption{
Mapping between allowed boson configurations, lozenge tilings and dimer coverings of a honeycomb lattice: a) shows the colored lozenge with a trimer-dimer pair and part of the honeycomb superlattice. b) shows an example configuration, where the lozenges arrange to form a dice lattice. The corresponding dimer covering on the honeycomb superlattice is indicated as well as the position of the dimers (fat lines). In c), we show two configurations that
contain a group of three lozenges, where the rotation of the bosons within the central hexagon leads to a flip between the two shown configurations. The notation for the two corresponding states in the quantum dimer model is shown in the bottom row.  
}
\label{fig:10_16__filling_with_dimers}
\end{figure}

In the classical limit, each valid configuration at this filling can be mapped to a lozenge tiling of the two-dimensional plane. Each lozenge is formed by eight lattice sites, and contains the  boson pattern shown in  \Fref{fig:10_16__filling_with_dimers}a, with a trimer-dimer pair such that one three-body vertex is introduced. 
A  boson covering of  filling $n=5/8$ and energy  $E_{pot}^{(5/8)}$ results whenever the full area of the lattice is covered with such lozenges. However, one has to introduce a parity constraint in mapping to the boson configuration: coloring  the lozenges as shown in \Fref{fig:10_16__filling_with_dimers}a, only sides of the lozenges of equal color are allowed to touch, in order not to introduce additional three-body repulsions, which would lead to a higher potential energy. 
Furthermore, it is well known that each lozenge tiling of the plane is dual to a hard-core dimer covering of a honeycomb lattice. 
This honeycomb lattice consists of hexagons that are a factor of two larger in linear extend than the hexagons of the underlying honeycomb lattice in the bosonic model.
Different lozenge tilings along with the corresponding dimer coverings and the underlying  boson patterns, are  shown  in \Fref{fig:10_16__filling_with_dimers}b and c. This mapping from bosonic configurations to dimer coverings allows to obtain the ground state entropy of the $n=5/8$ phase in the classical limit from that of closed-packed hard-core dimer coverings on the honeycomb superlattice, which equals $S=0.108$~\cite{moessner01} (the additional freedom of how the honeycomb superstructure is embedded onto the underlying lattice, and the two possible ways of coloring  the lozenges lead to a further, non-extensive factor of $3\times 2$ to the ground state degeneracy).
Among the various lozenge tilings we   find the dice lattice, shown in \Fref{fig:10_16__filling_with_dimers}b, which corresponds to a staggered dimer covering on the honeycomb superlattice. Other tilings contain a special group of three lozenges shown in \Fref{fig:10_16__filling_with_dimers}c. 
Rotating the bosons along the central hexagon in this structure, as shown in \Fref{fig:10_16__filling_with_dimers}c, leads to another allowed configuration, with the group of the three lozenges being rotated. In the dimer covering, this leads to a flip of the dimer pattern around the hexagon, as also seen in \Fref{fig:10_16__filling_with_dimers}c.
\begin{figure}
\centering
\includegraphics[width=300pt]{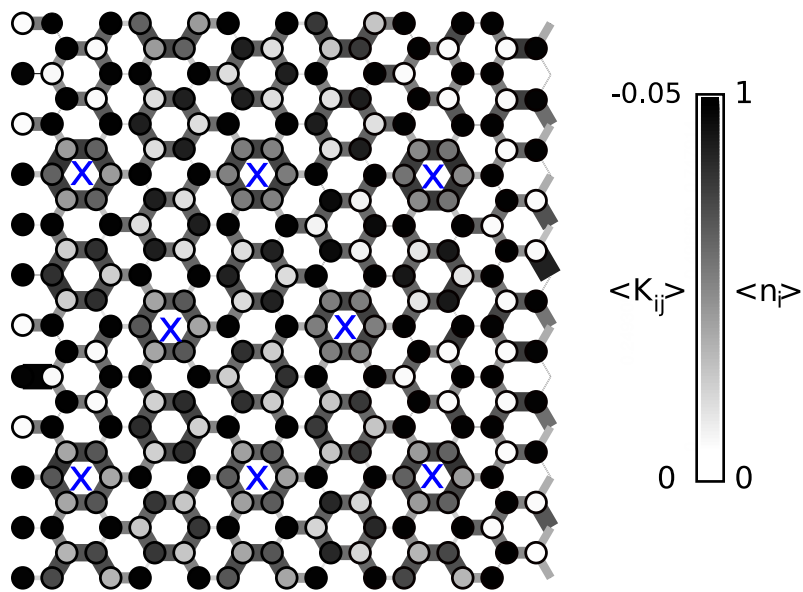}
\caption{
QMC data of the local density $\nn{n_i}$ (shading) and the kinetic energy density along the nearest-neighbor bonds $\langle K_ {ij}\rangle$ (line thickness and shading) for bosons on the honeycomb lattice 
in the $n=5/8$ VBC phase
at $V=0$, $t/W=0.2$, $\mu/W=3$, and a system with $L=12$ at $\beta=20$. Crosses indicate the most dominant plaquette resonances in this configuration. 
}
\label{fig:10_16__MC__filling_with_dimers}
\end{figure}

For finite $t$, 
a third-order boson hopping process 
corresponds to this local flip in the dimer configuration on the hexagonal plaquettes of the honeycomb superlattice, as illustrated in \Fref{fig:10_16__filling_with_dimers}c. 
Using degenerate perturbation theory in $t$, 
this local dimer-flip dynamics is described by a quantum dimer model with solely kinetic terms, 
\begin{equation}
H^\mathrm{QDM}\propto -t^3/W^2 \sum_{p}\left(  \ket{\alpha_p}\bra{\beta_p}+h.c. \right),
\end{equation}
and favors the formation of plaquette resonances as shown in \Fref{fig:10_16__filling_with_dimers}c.
Here, $\ket{\alpha_p}$ and $\ket{\beta_p}$ denote the two states on the hexagonal plaquette $p$ on the superlattice  in \Fref{fig:10_16__filling_with_dimers}c involved in the plaquette flip process. For finite values of $t$, the system  tries to maximize the number of plaquette resonances, and the ground-state degeneracy is  partially lifted~\cite{moessner01}. From the analysis of the quantum dimer model on the honeycomb lattice~\cite{moessner01}, we thus find that the ground state of the bosonic system at filling $n=5/8$  corresponds to the plaquette VBC phase of the quantum dimer model with purely kinetic terms. This phase is  characterized by resonances among  hexagons, and
we indeed observe a corresponding pattern in the QMC simulations: to exhibit this, we show in  \Fref{fig:10_16__MC__filling_with_dimers} the local density and the kinetic energy density on the nearest neighbor bonds for a representative point within the $n=5/8$ phase. The data is taken at $t/W=0.2$, outside the asymptotic regime where the perturbative derivation of the effective quantum dimer model strictly  holds. Nevertheless, we can identify in \Fref{fig:10_16__MC__filling_with_dimers}
a pattern that shares the structure of the plaquette VBC phase of the quantum dimer model, as indicated by  crosses on the most dominant hexagonal resonances.

\begin{figure}
\centering
\includegraphics[width=300pt]{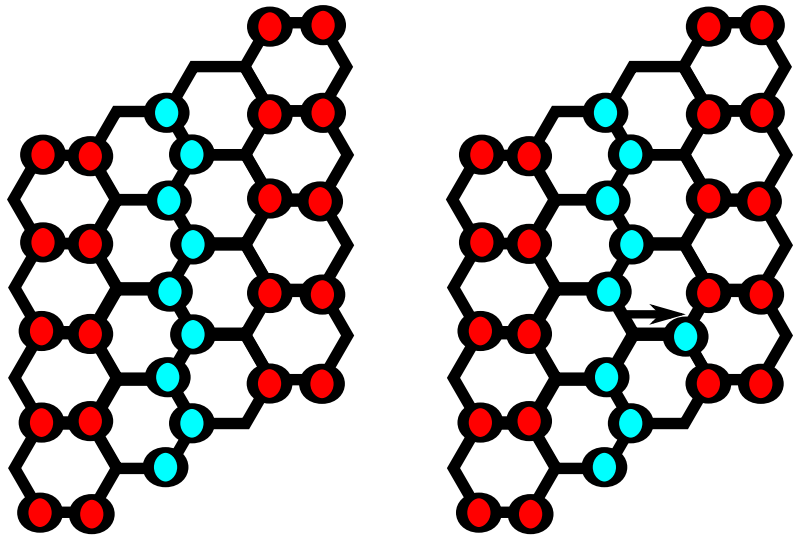}
\caption{
A configurations of bosons on the honeycomb lattice  at filling $n=2/3$, with potential energy $E_{pot}=-3W$ (left panel). Also indicated is one of the processes that leads to further patterns of the same density and potential energy (right panel). In the quantum model, such moves lead to resonances in the emerging $n=2/3$ phase.   
}
\label{fig:2_3_figure}
\end{figure}

\paragraph{The $n=2/3$ VBC Phase:}
Further increasing the chemical potential, 
the classical model exhibits a first-order phase transition at $\mu/W=5$ from the $n=5/8$ phase to a $n=3/4$ density plateau, with potential energy $E_{pot}^{(3/4)}=-3/4\mu+3/4W$, which (as well as $E_{pot}^{(5/8)}$) equals $-3W$ at $\mu/W=5$. This point in the phase diagram is highly degenerate; in particular, we find among the degenerate ground states also states with a filling of $n=2/3$ and  a potential energy  $E_{pot}^{(2/3)}=-2/3\mu+1/3W$ (equal to $-3W$ at $\mu/W=5$). One particular such state is shown in \Fref{fig:2_3_figure}. It consists of parallel strips of nearest neighbor boson pairs, separated by zig-zag chains of occupied sites. 
Other states of the same filling and energy can be obtained from this configuration upon allowing the bosons along the chains segments to move to a neighboring site. One such move is indicated in \Fref{fig:2_3_figure}.
Such processes however are not independent of each other: For example, if the  change indicated in \Fref{fig:2_3_figure} was made, the two bosons located at the next-nearest neighbors along the chain of the originally occupied site are blocked to perform a similar move, since that would lead to additional three-body repulsion terms, and thus a higher potential energy.
These changes in the configuration are thus blocking each other. 

\begin{figure}
\centering
\includegraphics[width=230pt]{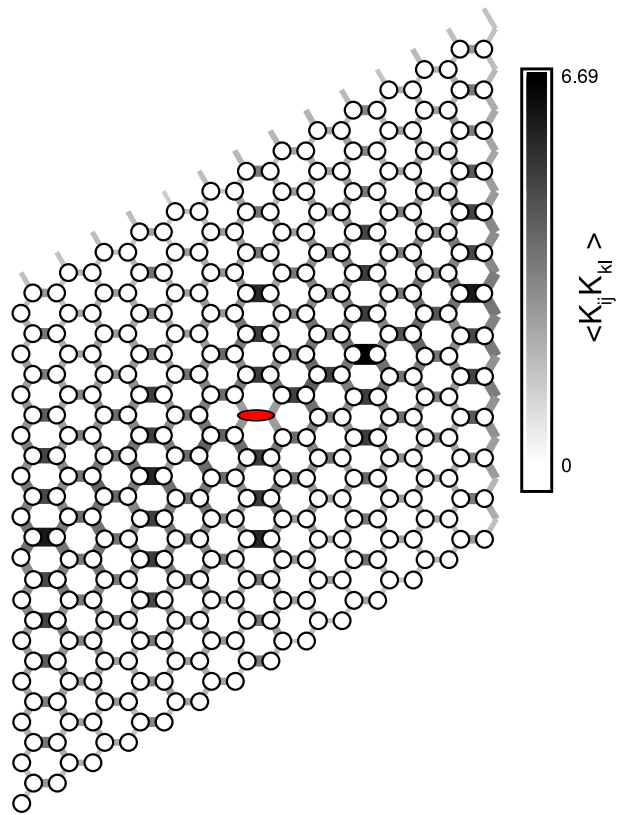}
\caption{
QMC data for the bond-bond correlations in the kinetic energy $\langle K_{ij} K_{kl} \rangle$ along the nearest-neighbor bonds for bosons on the honeycomb lattice 
in the $n=2/3$ VBC phase
at $V=0$, $t/W=0.3$, $\mu/W=5$, and a system with $L=12$ at $\beta=20$. The reference bond $\nn{ij}$ is indicated by the red ellipse.
}
\label{fig:2_3__MC__filling_and_corr}
\end{figure}

We find from the QMC simulation, that an extended phase of  filling $n=2/3$ gets selected via an order-by-disorder effect out of the degenerate ground state manifold in the classical limit at $\mu/W=5$: a new plateau of filling $n=2/3$ emerges in the quantum phase diagram of \Fref{fig:phase}. It vanishes in the classical limit, 
while upon increasing $t$, it 
extends  well into the regime  $\mu/W<5$. 
\begin{figure}
\centering
\includegraphics[width=300pt]{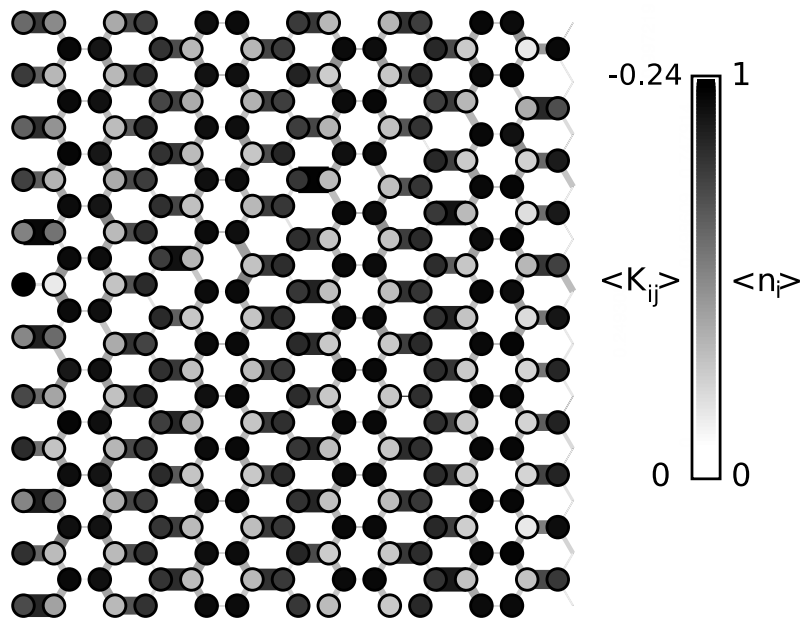}
\caption{
QMC data of the local density $\nn{n_i}$ (shading) and the kinetic energy density along the nearest-neighbor bonds $\langle K_ {ij}\rangle$ (line thickness and shading) for bosons on the honeycomb lattice 
in the $n=2/3$ VBC phase
at $V=0$, $t/W=0.3$, $\mu/W=5$, and a system with $L=12$ at $\beta=20$.
}
\label{fig:2_3__MC__filling_and_kin}
\end{figure}
In order to analyse the nature of this emerging phase, we show in 
\Fref{fig:2_3__MC__filling_and_kin} representative QMC data of the local  density and kinetic energy density within the $n=2/3$ phase. One  identifies a rigid backbone of parallel strips of boson pairs, where the local density takes on values close to unity. Furthermore, between these strips we find sites with  a reduced local density,  which are linked perpendicular to the stripes' direction by bonds with an enhanced kinetic energy density. 
Such behavior in both the density and the kinetic energy distribution is in accord with bond resonances,  induced by the  processes illustrated in 
\Fref{fig:2_3_figure}. 
The opening of the $n=2/3$ plateau for finite $t/W$ can thus be understood given the large local kinetic energy that the system gains, and which outweighs the penalty in potential energy. Since in the classical limit the kinetic energy contribution vanishes, the  potential energy penalty leads to the disappearance of this phase.
In \Fref{fig:2_3__MC__filling_and_corr}, we furthermore show the bond-bond correlations in the kinetic energy for the $n=2/3$ phase. 
While the correlations decay quickly in the direction parallel to the strips due to the previously mentioned blocking effect, they sustain over a rather wide range perpendicular to the stripe direction. This underlines the robustness of this emerging $n=2/3$ phase in the quantum regime. 

\paragraph{The $n=3/4$ Solid Phase:}

\begin{figure}
\centering
\includegraphics[width=300pt]{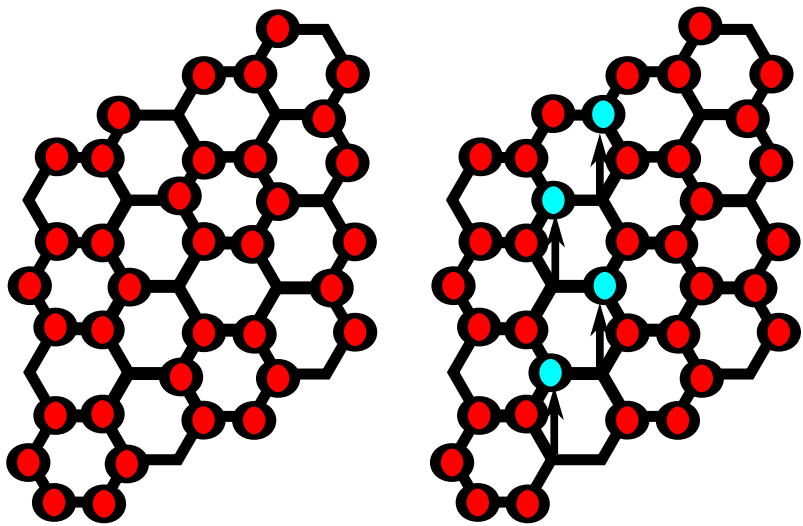}
\caption{
Two classical configurations  of filling  $n=3/4$. The state in the left panel consists of a regular superlattice of fully occupied hexagons. To obtain the state in the right panel, the particle colored blue have been shifted upwards (as indicated by arrows) with respect to their original positions in the left panel along two parallel lines. 
}
\label{fig:12_16__global_move}
\end{figure}

Increasing the chemical potential beyond $\mu/W=5$ , the classical system enters a phase of filling $n=3/4$, that concludes into the fully occupied state for $\mu/W>9+3t/W$. A particular classical state of this filling and a potential energy of $E_{pot}^{(3/4)}=-3/4\mu+3/4W$ consists of a regular superlattice of fully occupied hexagons shown in
the left panel of \Fref{fig:12_16__global_move}.
Additional states of the same filling and potential energy can be constructed by applying \textit{global} moves along parallel lines throughout the system, as shown in Figure \ref{fig:12_16__global_move}.
Such global moves shift all bosons on two parallel lines along one sublattice. The classical ground state entropy of this phase can be calculated by counting the number of states $W$ obtained by  such moves. 
As the number of lines to perform such moves is proportional to  the linear system size $L$, the  entropy per sites $S=(\ln W)/(2L^2) \propto 1/L$   scales to zero in the thermodynamic limit.

\begin{figure}
\centering
\includegraphics[width=300pt]{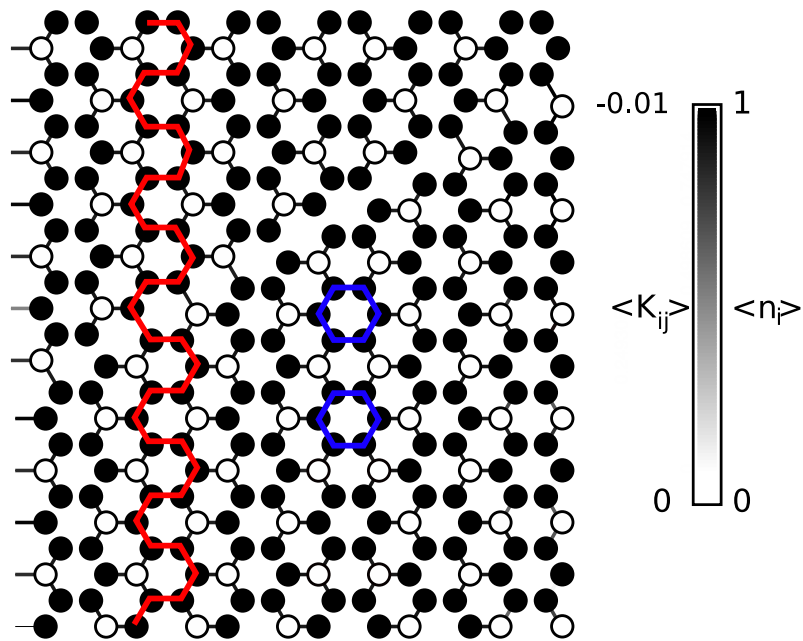}
\caption{
QMC data of the local density $\nn{n_i}$ (shading) and the kinetic energy density along the nearest-neighbor bonds $\langle K_ {ij}\rangle$ (line shading) for bosons on the honeycomb lattice 
in the $n=3/4$ solid phase
at $V=0$, $t/W=0.3$, $\mu/W=5$, and a system with $L=12$ at $\beta=20$. 
Two of the fully occupied hexagons observed in this structure are highlighted as well as a chain-like structure like the 
one  in Figure \ref{fig:12_16__global_move}.
}
\label{fig:12_16_kin}
\end{figure}

Because these degenerate ground states are connected not by local moves, but by global displacements, we do not find resonant structures  for finite $t$. Due to the global nature of the moves that relate the various classical ground states, we expect the QMC algorithm to generate states that are members of this rigid  manifold. Indeed, from the QMC simulations, we obtain 
density patterns that show characteristics of the deformed superlattice of fully occupied hexagons shown in \Fref{fig:12_16__global_move}.
A representative result from the QMC simulations is shown in Figure~\ref{fig:12_16_kin}. 
The classical degeneracy thus appears not to be lifted in the quantum regime.

\subsection{VBC-VBC Quantum Phase Transitions}

After having described the  VBC phases appearing in the presence of three-body repulsions on the honeycomb lattice, we next  turn to  the quantum phase transitions between these incompressible phases.
An exploration of the transition regions is challenging for the QMC algorithm, because  neighboring VBC states are rather close in energy, with competing potential and kinetic contributions of similar size.
To illustrate this issue, we consider a numerical example of the relevant energy scales. The potential energy difference between the two phases at $9/16$ and $5/8$ filling is $\Delta E_\mathrm{pot} = -1/16 \mu + 1/8 W$, which equals $-0.0625W$
near $\mu/W=3$. The kinetic energy difference between the two VBC phases for a given value of e.g. $t/W=0.3$ is obtained from the QMC simulations as 
$\Delta E_\mathrm{kin}(t/W=0.3)\approx 0.025W$, and is thus of similar size.
We  find that the competition between these two similar energy scales leads to an extended transition region between the VBC phases, that gives rise to 
an apparent continuous increase in the density, as seen e.g. in the inset of \Fref{fig:phase}. We explicitly checked that such behavior occurs also for temperatures as slow as $T/W=0.05$ and $T/W=0.01$, i.e. well below the above energy scales. 
Furthermore, we find strong algorithmic hysteresis effects upon varying the chemical potential at fixed $t/W$ through the  transition region within a single  simulation. 
\begin{figure}
\centering
\includegraphics[width=300pt]{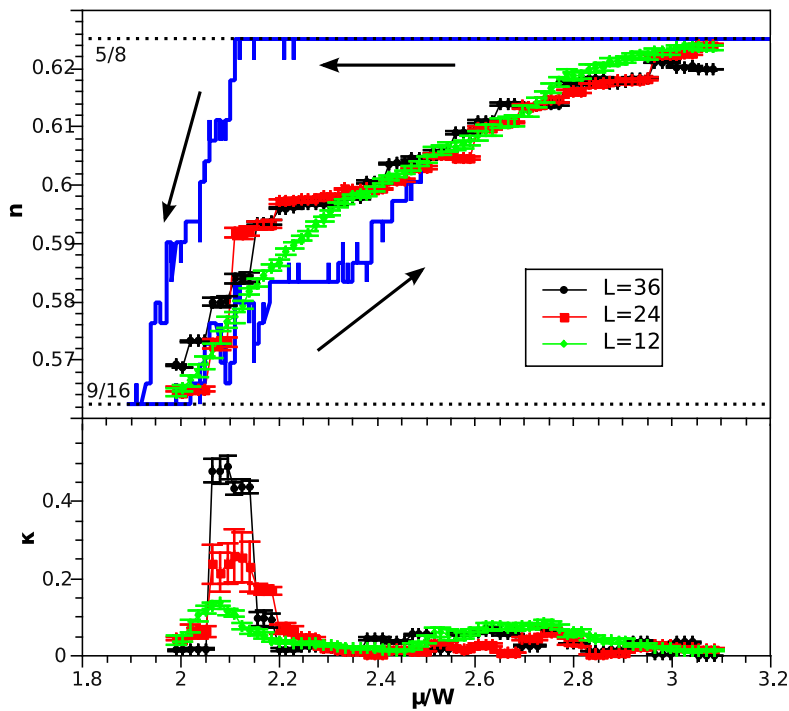}
\caption{
Results of a hysteresis study between the $9/16$ and $5/8$ VBC phases (solid blue line) for a system size $L=12$, and results from quantum parallel tempering simulations for systems of  sizes $L=12,24$, and $36$. The upper panel shows the 
filling $n$, and the lower panel the compressibility $\kappa$  obtained from the quantum parallel tempering calculations.
}
\label{fig:full_hyst}
\end{figure}
\Fref{fig:full_hyst} exemplifies this behavior: 
Starting within the $n=5/8$ phase and decreasing $\mu$, the density is stable until reaching $\mu/W \sim 2.1$, where it drops to $n=9/16$ rapidly. Starting from the $n=9/16$ phase and increasing $\mu$, the filling of $n=5/8$ is not
established even well inside the $n=9/16$ VBC region. This behavior exhibits a metastability of the
$n=9/16$ state; the algorithm cannot 
perform effectively the re-arrangements that are necessary in order to establish the structure of the $n=5/8$ phase, 
starting from a  configuration typical for the $n=9/16$ phase. 

We  employed quantum parallel tempering in $\mu/W$ in order to assess, if with the help of replica-exchanges  between neighboring values of $\mu/W$, the algorithm  overcomes such metastabilities. Results of these calculations are shown in \Fref{fig:full_hyst} for systems of different sizes for $V=0$ and $t/W=0.3$.  We find that
the obtained density $n$ mimics the behavior of the hysteresis curve. 
In particular, we observe an increase of $n$ within the range of $\mu/W$, where the upper hysteresis curve drops.
In addition, the compressibility $\kappa$ shows a peak within this region, that increases with system size. In the QMC simulations, the compressibility
$\kappa= \partial n/\partial \mu=\beta (\langle n^2\rangle - \langle n\rangle^2)$ is  obtained in terms of the density fluctuations. 
The density curve flattens for the larger values of $\mu/W$, and does not reach $5/8$ even at $\mu/W=3$. On the other hand, $n$ reaches the value  $9/16$ for $\mu/W>2$. Hence, the replicas are not able to establish the $n=5/8$ VBC structure, even though they  tunnel repeatedly throughout the extended parameter range. These observations are consistent with a  first-order quantum phase transition between the neighboring VBC phases.

However, one might be concerned, that the numerical difficulties could hide an intermediate phase that separates 
the VBC phases.
In one such 
scenario, an intermediate phase is not commensurate with our lattice sizes. When simulating lattices of varying sizes ($L$=12, 24, 25, 36), we did however not detect any new structures in the transition regime. Of course, we cannot exclude phases with very large unit cells from our finite size study. 
Another possibility would be a superfluid or even supersolid phase in the intermediate region, such as  observed in a similar model on the square lattice~\cite{schmidt08}. 
However, we can rule out a finite superfluid density in the transition region:
Using quantum parallel tempering in $t/W$ or driving a superfluid system (from large $t/W=0.5$) towards the transition region at low $t$,  resulted in $\rho_s$ always becoming zero for values of $t/W \lesssim 0.35$. 
A third possibility for an intermediate phase would be a VBC "emulsion" i.e., a metastable phase mixture with domain walls separating $9/16$ and $5/8$ VBC-like domains. While we observe in the QMC simulations bosonic structures that show features of both the $9/16$ and the $5/8$ phase, which would be expected from such an VBC "emulsion" scenario, these structures are also consistent 
with a first-order transition, given the algorithmic metastability. 
Thus, we consider a first-order VBC-VBC quantum phase transition the most conservative scenario consistent with the QMC data and the apparent difficulty of the QMC algorithm in this parameter regime. 
While we cannot determine the precise location of the quantum phase transition,  we take the peak position of the compressibility as an estimate.
This is reflected in the phase diagram in \Fref{fig:phase}: The points denote the region where the phase transition  occurs according to our  estimate. Since we are however not able to provide the exact position, but  an interval, we denote this interval by an errorbar. 
We note, that the above discussion applies likewise to the other inter-plateau transitions.

\subsection{Classical Limit ($t=0$)}

\paragraph{Finite Cluster Studies:}

In order to analyse the ground state phase diagram of the present model, it is useful to consider also the classical limit $t=0$  of the quantum Hamiltonian.  In the particular case  $V=0$, the model reduces to the most simple classical model on the honeycomb lattice with extended three-body interactions. We are not aware of any previous numerical or even exact results on this statistical physics model. While one can construct states that appear in the classical limit from analysing boson configurations from the QMC simulations, we  wanted to check the implications for the classical limit using independent methods. 
For this purpose,  we  solved the classical model on a finite hexagonal cluster of linear system size $L=4$ exactly. Doing so, we confirm the extends of the densities plateaus as described above, including the absence of a $n=2/3$ plateau in the classical limit. 
However, this system size suffers from the particular problem, that among the various  states of the $5/8$ plateau, only the
bosonic state  corresponding to the 
dice lattice lozenge tiling (c.f.  \Fref{fig:10_16__filling_with_dimers}b) matches onto this finite cluster.
While this particular tiling provides a valid boson covering in the classical limit, this restriction shadows the huge degeneracy of this phase. Namely, on the $L=4$ cluster,  no configuration can be realized, that contains the group of three lozenges shown in \Fref{fig:10_16__filling_with_dimers}c.
In order to be commensurate with such  coverings, the linear system size must be an integer multiple of 6. All other phases however are commensurate with this system size, and we can exclude different boson patterns than those described above for up to $32$ sites exactly.

In order to check on larger clusters, whether the classical predictions for the bosonic structures are correct, we tried to employ classical Monte Carlo simulations. However, they suffer from dynamical freezing at the low temperatures required to explore the plateau structure. In that respect, the QMC simulations perform  more efficient, and we thus used the QMC simulations in order to generate configurations of the classical model (this is possible within the SSE framework since on each propagation level one obtains a configuration of the classical model). While we do of course not sample these classical configurations with the correct statistical weight, we can nevertheless check if the various classical configuration obtained this way are consistent with our effective descriptions. 
In particular, we verified  that each  classical configuration of density $n=5/8$ and potential energy $E_{pot}=E_{pot}^{(5/8)}$ indeed complies with the construction given in Section 2.2 in terms of lozenge tilings. 
We did not observe any classical configuration with the right density and potential energy, that would violate this construction. We are thus confident, that our understanding of the classical phases is correct. 

\paragraph{Tensor Network Renormalization Group:}
Recently, 
an interesting novel approach to study thermodynamic properties of classical statistical models has been  proposed. It is based on a tensor network representation of the partition function, and evaluates it directly in the thermodynamic limit using a renormalization procedure in the tensor network decomposition. For details about this method, we refer to 
the original paper by M. Levin and C. P. Nave \cite{levin07}, as well as   recent works applying this approach to the Ising-model on the triangular \cite{hinczewski08} and the Shastry-Sutherland lattice \cite{chang09}. Since this method performed well in these cases, we tried to apply it also to our model of nearest-neighbor three-body repulsions on the honeycomb lattice. In fact, in the original publication, the method was described explicitly for tensor network models on the honeycomb lattice. We found, that one can indeed express the partition function of our model in terms of a tensor network. For this purpose, 
one specifies two cyclically symmetric tensors $T^A_{ijk}$ and $T^B_{ijk}$, with indices $i,j,k$ running from $1$ up to a finite integer $D$, where each index  corresponds to  a degree of freedom $i=1,...,D$ on the bonds of the honeycomb lattice. 
Due to the bipartiteness of the honeycomb lattice, three bonds  meet  at a site of sublattice $A$ or $B$. The tensor $T^A_{ijk}$ is  assigned to each site of the honeycomb lattice within sublattice $A$, and likewise the tensor  $T^B_{ijk}$
to those of sublattice B. 
The partition function of a tensor network model on the honeycomb lattice is then given as 
\begin{equation}
 Z=\sum_{i,j,k,...=1}^D T^A_{ijk}T^B_{ilm}T^B_{jnp}T^B_{kqr}...,
\end{equation}
i.e. $Z$ is obtained from the product of all the tensors, contracting pairs of indices for each bond of the honeycomb lattice.
In order for $Z$ to match the partition function of the bosonic model (in the classical limit $t=0$), the tensors $T^A_{ijk}$ and $T^B_{ijk}$ can be chosen as given  in Table 1, with the dimension $D=4$. Here, we considered the general case, where both $W$ and $V$ are finite.
One verifies easily, that with these particular tensors, $Z$ indeed recovers the partition function of the classical particle model.
\begin{table} 
\begin{center}
\begin{tabular}{| c | c | c || l | l |}
  \hline
  i&j&k&$T^{A}_{ijk}$&$T^{B}_{ijk}$ \\
  \hline
  1 & 1 & 1     & 1                     & 1 \\
  1 & 2 & 2     & 1                     & 0 \\
  1 & 1 & 3     & 0                     & 1 \\
  2 & 2 & 2     & 0                     & $\exp(\beta\mu)$ \\
  3 & 3 & 3     & $\exp(\beta\mu)$     & 0 \\
  2 & 2 & 1     & 1                     & 0 \\
  3 & 3 & 1     & 0                     & 1 \\
  3 & 3 & 4     & $\exp(-\beta (-\mu+V/2))$& 0 \\
  2 & 2 & 4     & 0                     & $\exp(-\beta (-\mu+V/2))$ \\
  3 & 4 & 4     & $\exp(-\beta (-\mu+V+W))$& 0 \\
  4 & 4 & 2     & 0                     & $\exp(-\beta (-\mu+V+W))$ \\
  2 & 2 & 2     & 1                     & 0 \\
  3 & 3 & 3     & 0                     & 1 \\
  4 & 4 & 4     & $\exp(-\beta(-\mu+3/2V+3W))$ & $\exp(-\beta(-\mu+3/2V+3W)$ \\
  \hline
\end{tabular}
\label{ }
\caption{Non-zero tensor elements $T^{A}_{ijk}$ for links $i,j,k$ around a site of sublattice $A$, and
$T^{B}_{ijk}$ for links $i,j,k$ around a site of sublattice $B$, respectively. All other finite
tensor elements are obtained from the shown ones via cyclic permutation of the indices $i,j,k$.
}
\end{center}
\end{table}
Once a representation of the classical model in terms of a tensor network has been obtained, we can proceed to perform the renormalization procedure to this tensor network. Doing so, involves an approximation, since within each renormalization step the dimension of the renormalized tensor network is truncated to a fixed maximum dimension $D_{max}>D$.  $D_{max}$ is thus a regularization parameter of the algorithm.From the approximatively evaluated partition function, one obtains the free energy $F=-\frac{T}{N}\ln{(Z)}$ in the thermodynamic limit, from with the density  $n$ is  calculated by taking numerical derivatives of $F$ with respect to $\mu$. Doing so for sufficiently low temperatures $T$, we  obtain the low-$T$ phase diagram from resolving the density plateau structure. For sufficiently large $D_{max}$, the numerical data eventually converges to the final result. Within the tensor network renormalization procedure, one performs a singular value decomposition of a $(D_{max})^2 \times (D_{max})^2$ matrix of contracted tensors, from which the renormalized tensors are constructed. After each step the tensors have be to be normalized such that all tensor elements are smaller or equal to unity to avoid overflows. Since this procedure is iterated many times, one needs to implement all matrix computations using high-precision floating point arithmetics (e.g. the initial non-zero tensor elements differ in about 18 orders of magnitude from for $\mu/W=2$ and $\beta=20/W$ at $V=0$.). In Ref.~\cite{hinczewski08}, quadruple precision was found appropriate for an  Ising model. Using a customized version of LAPACK~\cite{laug} with 128 bit reals (floating point precision of about $10^{-34}$), we find that for the current model this precision was  still not sufficient in the relevant part of the phase diagram.  While the  method   performs for the case of purely two-body interactions (i.e. for $W=0$),  the tensor iteration procedure does not converge for $\beta\gtrsim W$, once $W$ dominates and the chemical potential reaches beyond $\mu/W\approx 1$. (For $\beta<W$, the method converged and the results could be verified by comparing to Monte Carlo simulations. However, for such high temperatures, the plateau structure is thermally smoothed out, thus providing no information about the ground state properties, which we are after.)
\begin{figure}
\centering
\includegraphics[width=250pt,]{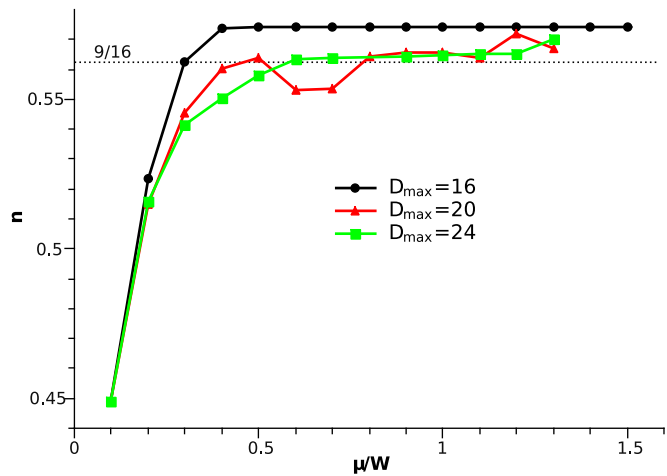}
\caption{
Filling $n$ in the classical limit at $V=0$, as a function of $\mu/W$ obtained from the tensor network renormalization group approach at $\beta=10$. The curves correspond to different values of $D_{max}$.
}
\label{fig:tnrg_3}
\end{figure}
As an example, we show in Fig.~\ref{fig:tnrg_3} the density curves that we obtained at $\beta=10$.
In agreement with our previous analysis, we still find that the system enters a $n=9/16$ plateau. However,  we cannot explore the full phase diagram using this method. We  trace the failure of the tensor network renormalization group procedure
to the fact, that the tensors for larger values of $\mu/W$ span many orders of magnitude due to the exponential dependence of the Boltzmann factor in the tensor elements. 
Thus a broad range of values is required in order to account for the physics of the model. However, due to the accumulation of round-off errors in the computation of the renormalized tensor network (which includes a large number of additions and multiplications), one loses the required precision in a numerical implementation of the algorithm
after a small number of iteration steps. This problem appears to be a generic drawback of the tensor network renormalization group method, which we suffer from because
of the broad range of magnitudes that our system implies in the tensor network representation. For a model with  two-body interactions only (i.e. at $W=0$), the range of values is significantly reduced, and in this case we could indeed
recover the (well known) behavior of the model, which is equivalent to the ferromagnetic Ising model on the honeycomb lattice.
\begin{figure}
\centering
\includegraphics[width=200pt,angle=270]{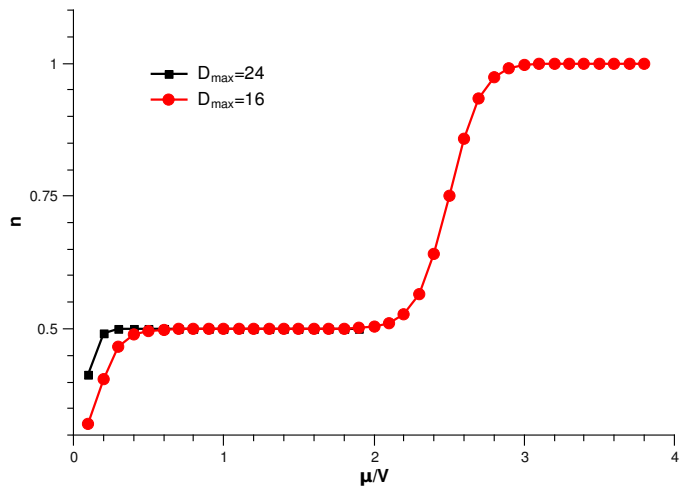}
\caption{
Filling $n$ in the classical limit ($t=0$) at $W=0$, as a function of $\mu/V$ obtained from the tensor network renormalization group approach at $\beta=10$. The  curves correspond to different values of $D_{max}$.
}
\label{fig:tnrg_2}
\end{figure}
As an example, we show in Fig.~\ref{fig:tnrg_2} the filling $n$ as a function of $\mu/V$ at $W=0$  and $\beta/W=10$,
with a pronounced plateau at $n=1/2$ emerging.

\section{Two-body Interactions}
Thus far, we considered mainly the case of  purely three-body repulsions, and explored the phases of the bosonic model in that parameter regime. However, from the derivation of the extended Hubbard model for ultra-cold polar molecules in Ref. \cite{buechler07} it is clear,  that  two-body interactions will be at least of the same strength as the three-body terms (the claim in Ref.~\cite{schmidt08} , that the interactions are solely of three-body type thus appears in  contrast to the results in Ref.~\cite{buechler07}). Hence, it is important to assess the influence of two-body terms on the physics of such  models. Starting with the nearest-neighbour three-body term $W$, the next important interaction term  to be considered is the nearest-neighbor two-body repulsion $V$. 

The phase diagram for hard-core bosons with solely nearest-neighbor two-body interactions ($W=0$, $V\ne 0$) features a half-filled checkerboard solid for small values of $t/V<0.5$, surrounded by a superfluid phase, without any supersolid phases present in the quantum phase diagram~\cite{wessel07b}. In the current setup, we recover this checkerboard solid at large values of $V/W$. An important question is,  whether the other phases discussed in Section 2 are  stable towards the relevant parameter regime $V \gtrsim W$, and if new phases appear from the competition between two- and three-body interactions. 

\paragraph{Cascaded Transition:}
To make predictions about the impact of two-body interactions, we first consider the $n=9/16$ phase in the classical limit. It consists of triangles with an edge length corresponding to the size of four hexagons (c.f. \Fref{fig:9_16}). No three-body vertices are present in this state, but the structure has neighboring bosons along the edges of the triangles. For finite $V$, these boson pairs result in a potential energy penalty, which tends to destabilize the structure.

\begin{figure}
\centering
\includegraphics[width=230pt]{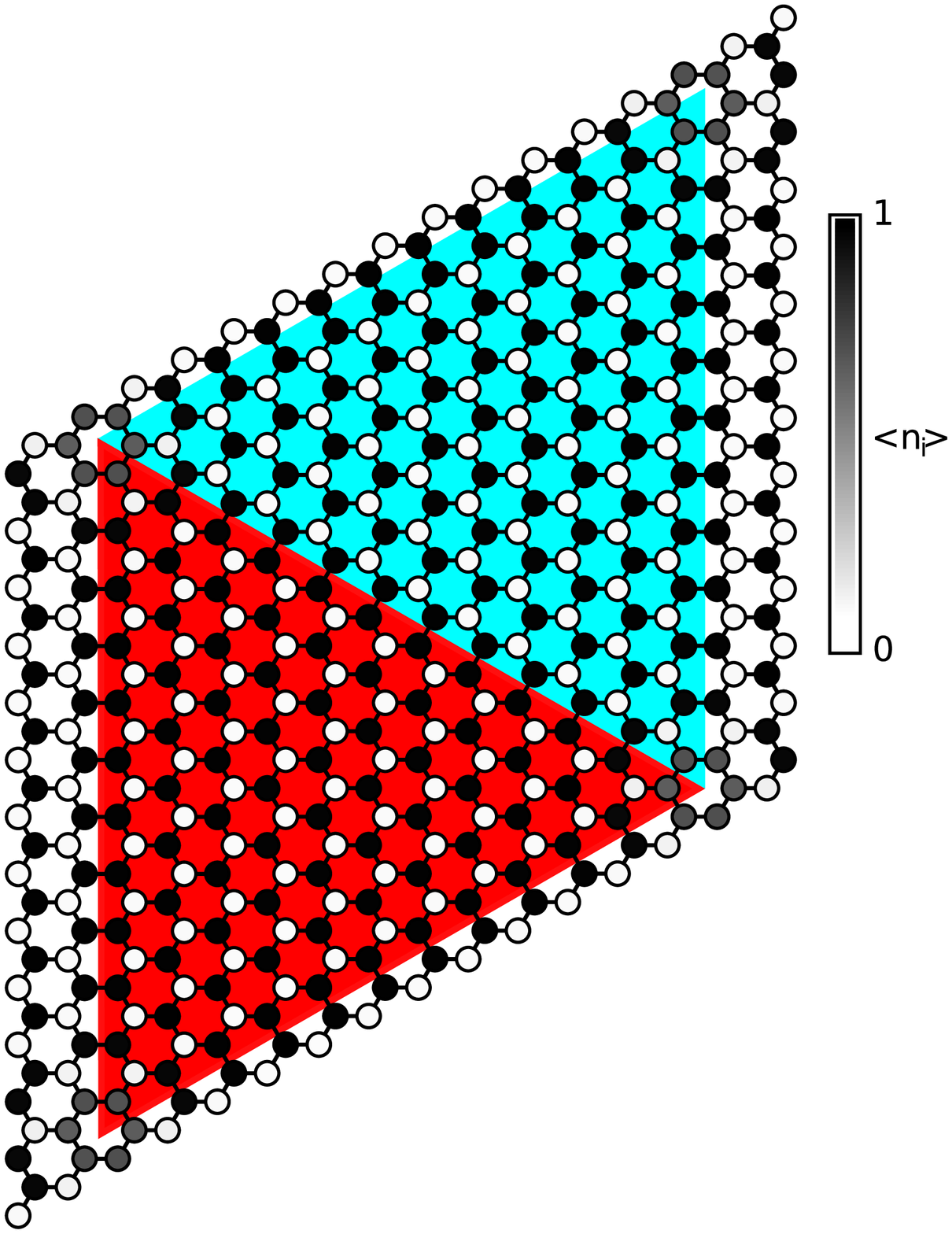}
\caption{
QMC data for the local density $\nn{n_i}$ (shading) for $\mu/W=4$, $t/W=0.3$ and $V/W=1.25$. The colored triangles  
highlight the unit cell of  the superstructure with $x=12$ found in the density plateau at filling $n=77/144$. 
}
\label{fig:x12}
\end{figure}

One possibility would be, that there is a direct transition from the $n=9/16$ phase to the $1/2$ solid upon increasing $V/W$. However, we find that one can construct  intermediate states with densities $1/2 <n <9/16$, that are energetically preferred for certain ranges of $V/W$. We obtain such state, upon 
generalizing the classical $n=9/16$ state, where  checkerboarded (half-filled) triangles are separated by domain walls: we  now consider the size of these triangular domains to vary.
Namely, we
consider a honeycomb lattice and cover it with equilateral triangles of edge length $x$ (in units of the size of one hexagon). Each triangle thus covers $x^2$ lattice sites (\Fref{fig:x12} shows QMC data for the local density that corresponds to a configuration with $x=12$). A staggered filling with bosons  yields a lattice filling of $n_\triangle(x)=x(x+1)/2-1$. The boson pairs along the boundaries of the triangles cost an energy of $P_\triangle=V(3x-4)/2$, and we obtain the potential energy per lattice site as 
\begin{equation}
E_\triangle(x)=-\frac{\mu}{x^2} n_\triangle(x) + \frac{1}{x^2} P(x)_\triangle V. 
\end{equation}
Minimizing the energy with respect to $x$ gives $x=4$ for $V=0$ (we thus recover the states at filling $9/16$) and $x\rightarrow \infty$ as $V/W\rightarrow \infty$ (we converge to the half-filled checkerboard state). 
The number of resonances in a system with $N$ sites equals $N/(2x^2)$, resulting in a macroscopic ground state 
degeneracy of entropy $S/N=\ln(3)/(2x^2)$ for these solid phases.

\begin{figure}
\centering
\includegraphics[width=350pt]{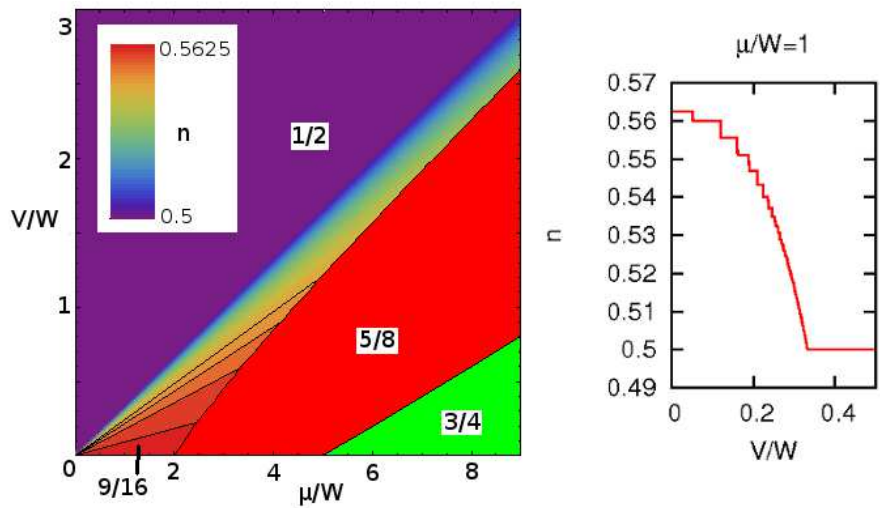}
\caption{Classical phase diagram for finite $V$ and $W$ in terms of $\mu/W$ and $V/W$ (left panel). Colors indicate the filling of the corresponding phases. The $n=5/8$ and $n=3/4$ plateaus extend towards finite values of $V/W$. The color gradient denotes the filling in the cascade, varying from $9/16$ to $1/2$. The phase boundaries of the larger phases are marked by black lines. The right panel shows the filling $n$ as a function of $V/W$ in the classical limit at $\mu/W=1$.}
\label{fig:twobody_classical}
\end{figure}

In order to derive the ground-state phase diagram based on this construction, we minimize the energy for fixed $\mu/W$ and $V/W$ as a function of $x$. For $\mu/W\ge 2$ other competing states are the $n=5/8$ and for $\mu/W \ge 5$ the $n=3/4$  phase described in Section 2. We thus optimise the energy among these various states in order to determine the phase boundaries. 
The phase diagram obtained this way is shown in the left panel of in \Fref{fig:twobody_classical}. 
We find that in addition to the previously established phases, the system exhibits a whole cascade  of new solid structures (with $x$ running from $4$ to infinity) that appear upon increasing the value of $V/W$. As an example, we show in the right panel of \Fref{fig:twobody_classical} the filling as a function of $V/W$ for a fixed value of $\mu/W=1$. Starting from  the $n=9/16$ plateau, a cascade of transitions eventually leads for $V/W>0.35$ to the staggered plateau of filling $1/2$. This cascade of plateau phases forms an incomplete devil's staircase, induced by the competing nature of the three- and two-body repulsion terms. 

\begin{figure}
\centering
\includegraphics[width=250pt]{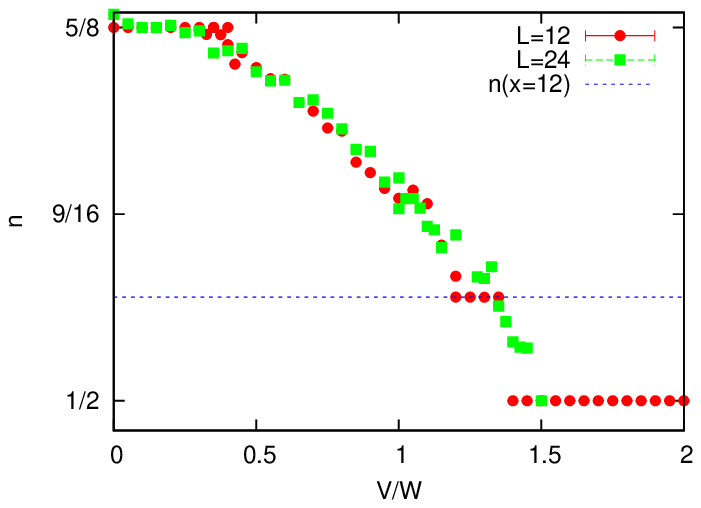} 
\includegraphics[width=250pt]{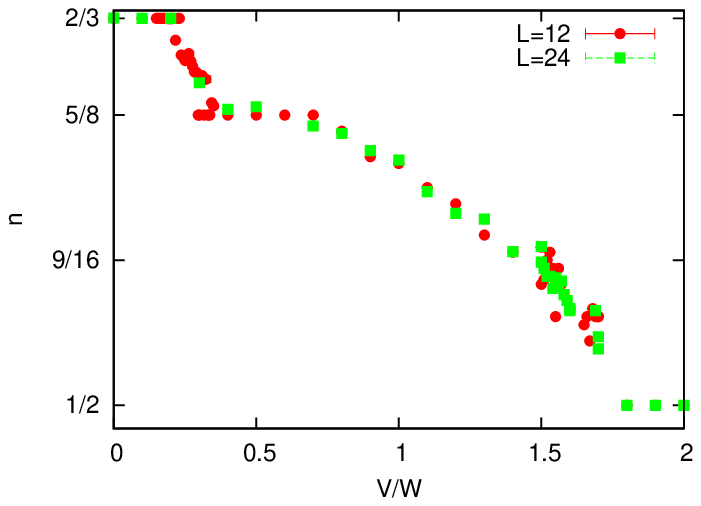}
\includegraphics[width=250pt]{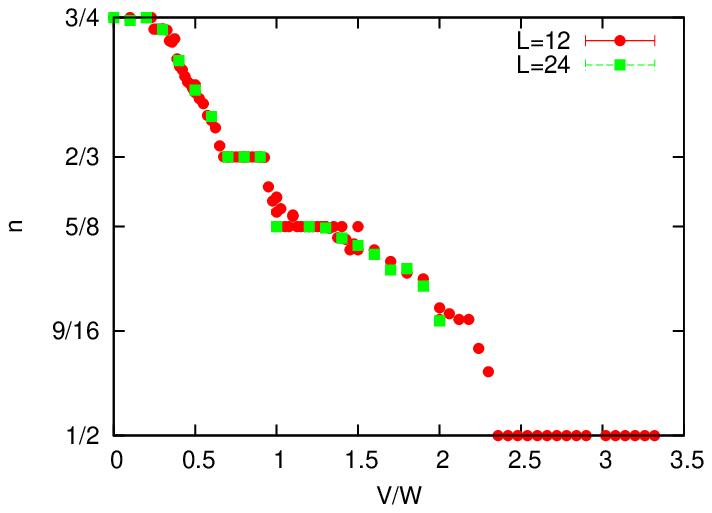}
\caption{
QMC data of the filling $n$ as a function of $V/W$ for $t/W=0.3$ and $\mu/W=4,\,5$ and $7$ (from top to bottom) for $L=12$ and $24$ at $\beta=20$. \textbf{Top:} For $V/W \approx 1.2$ a plateau at $n=77/144$ (corresponding to $x=12$) appears, with a direct transition to  the half-filled solid at $V/W \approx 1.4$. \textbf{Middle:} The $2/3$ density plateau is stable towards finite $V/W$ and decays into the $5/8$ plateau at $V/W \approx 0.3$. \textbf{Bottom:} For large enough $\mu/W$ one can also see the finite extend of the $3/4$ plateau. 
}
\label{fig:twobody_filling}
\end{figure}

\paragraph{Numerical Results:}
For finite $t>0$, we expect deviations from the classical cascade structure, since the number of hexagonal resonances that appear for finite $t$ 
decreases  as $1/x^2$ with increasing $x$. The system could thus skip some of the higher-$x$ plateaus because of the stabilization of the low-$x$ phases by their larger kinetic energy. This can result in  entering  the staggered phase after only a finite number of intermediate plateaus. Besides, quantum fluctuations stabilize the $n=2/3$ phase and we have to account for the relevance of this new phase on the  overall phase diagram. However, it is  difficult to resolve most of the new solid structures numerically  because they are not commensurate with our lattice sizes. Moreover, the differences in energy and filling for states of neighboring values of $x$ are small, and the plateaus  rather narrow already in the classical limit. We nevertheless obtain evidence from the QMC simulations for (i) a non-direct transition to the staggered solid upon turning on a finite two-body repulsion $V$, and (ii) the stabilization of new VBC phases. 
For this purpose, \Fref{fig:twobody_filling} shows QMC data for the $V$-dependence of the filling $n$ for $\mu/W=4,~5$ and $7$ respectively, obtained at $t/W=0.3$. There one clearly identifies a plateau corresponding to the $x=12$ structure (of filling $n=77/144$), commensurate with our lattice sizes of $L=12$ and $24$. In \Fref{fig:x12}, we show a representative boson covering obtained by  QMC for $\mu/W=4$ at $V/W=1.25$, which is in perfect agreement with the $x=12$ structure introduced above and the formation of hexagonal resonances as in the $n=5/8$ phase. 
From \Fref{fig:twobody_filling}, we find that the $n=2/3$ phase is stable for small finite values of $V/W<0.2$, before a transition takes place towards the $n=5/8$ density plateau. Similarly, we find from \Fref{fig:twobody_filling}, that the $n=3/4$ phase remains stable for small values of $V$. At larger values of $V/W$ we clearly resolve the $n=2/3$ and the $n=5/8$ plateau.
While we thus obtain  evidence for a cascaded transition to the checkerboard solid, we were not in a position to fully resolve this transitions within our finite size QMC simulations. We hence  did not attempt to comply a full phase diagram for finite $V$ in the quantum case, but  considered the above  set of representative cuts through the parameter space. 
However, considering the relevant interaction range 
$V\gtrsim W$~\cite{buechler07}, we still find that in particular the $n=5/8$ VBC phase can be realized for such realistic 
values of the ratio between three- and two-body interaction terms.

\section{Conclusion}
We  studied a model of hard-core bosons with strong three-body repulsions on the honeycomb lattice using quantum Monte Carlo simulations,  exact cluster analysis and the tensor-network renormalization group approach. The system's ground state phase diagram  exhibits besides a superfluid region several complex valence bond crystal phases at fractional fillings $9/16$, $5/8$, $3/2$ and $3/4$. The obtained quantum phase transitions between neighboring valence bond crystal phases are consistent with first-order transitions, given mild energy differences in both the potential and the kinetic energy sectors.  
With regard to a possible experimental realization based on cold polar molecules in appropriately tuned external electric and microwave fields, 
we included in addition to the three-body repulsions also nearest neighbour two-body interactions,  which in a realistic set-up are at least of the same strength as 
the three-body 
interactions. Considering the competition between the different interaction terms, 
we obtained a cascade of intermediate incompressible plateaus as the two-body interaction strength  increases. Furthermore, we find that the valence bond crystal 
phase of filling $5/8$ with an effective low-energy description in terms of a quantum dimer model remains  accessible well  within the  reachable 
parameter regime. 
A further step towards  modeling cold polar molecules on the honeycomb lattice would be to take into account the full long-ranged nature 
of both the three-body and two-body interactions~\cite{buechler07}. Treating the longer ranged contributions appropriately appears to  require the use of 
different calculational techniques, since already the leading contributions to both interaction sectors provided a challenge for the quantum Monte Carlo 
approach. Based on our current results, one could expect that  longer-ranged interactions stabilize additional solid phases with  large unit cells, at least in the classical regime. For finite hopping strengths, residual entropies of the classical states would be lifted, eventually leading to the 
emergence of complex resonating structures similar to those described above. Whether other exotic states  e.g. with  topological order can indeed be stabilized in three-body extended Bose-Hubbard models~\cite{buechler07}, thus far appears  open to future investigations.

\section*{Acknowledgements}
We should like to thank K. P. Schmidt, A. L\"auchli, R. Moessner, and A. Muramatsu for interesting discussions, and acknowledge the allocation of CPU time on the high performance computers at HLRS Stuttgart and NIC J\"ulich, where the numerical calculations have been performed. L. B. furthermore acknowledges support from the Studienstiftung des Deutschen Volkes, and HP. B. and S. W. from the DFG within the SFB/TRR 21.

\end{document}